\documentclass[preprintnumbers,floatfix,letterpaper,aps,prd, nofootinbib,longbibliography,twocolumn]{revtex4-2}
\usepackage{graphicx}
\usepackage{dcolumn}
\usepackage{bm}
\usepackage{hyperref}
\usepackage[utf8]{inputenc}
\usepackage{booktabs}
\usepackage{multirow}
\usepackage{orcidlink}
\usepackage{etoolbox}

\begin{document}

\title{Tests of general relativity at the fourth post-Newtonian order with GW230627 and GW250114}

\author{Xi-Min Liang${}^{a, b}$ \orcidlink{0009-0004-7056-5418}}
%\email{lmliang@zjut.edu.cn}

\author{Yuan-Zhu Wang${}^{a, b}$ \orcidlink{0000-0001-9626-9319}}
%\email{vamdrew@zjut.edu.cn}

\author{Tao Zhu${}^{a, b}$ \orcidlink{0000-0003-2286-9009}}
\email{Corresponding author: zhut05@zjut.edu.cn}

\author{Wen Zhao${}^{c, d}$ \orcidlink{0000-0002-1330-2329}}
%\email{wzhao7@ustc.edu.cn}

\author{Xin Zhang${}^{e, f, g}$ \orcidlink{0000-0002-6029-1933}}
%\email{zhangxin@mail.neu.edu.cn}

\affiliation{
${}^{a}$Institute for Theoretical Physics \& Cosmology, Zhejiang University of Technology, Hangzhou, 310023, China\\
${}^{b}$United Center for Gravitational Wave Physics, Zhejiang University of Technology, Hangzhou, 310023, China\\
${}^{c}$CAS Key Laboratory for Research in Galaxies and Cosmology, Department of Astronomy, University of Science and Technology of China, Hefei 230026, China \\
${}^{d}$School of Astronomy and Space Sciences, University of Science and Technology of China, Hefei, 230026, China\\
${}^{e}$ Key Laboratory of Cosmology and Astrophysics (Liaoning), College of Sciences, Northeastern University, Shenyang 110819, China\\
${}^{f}$Key Laboratory of Data Analytics and Optimization for Smart Industry (Ministry of Education), Northeastern University, Shenyang 110819, China\\
${}^{g}$National Frontiers Science Center for Industrial Intelligence and Systems Optimization, Northeastern University, Shenyang 110819, China}

\date{\today}

\begin{abstract}

Gravitational wave (GW) observations provide an unprecedented laboratory for testing general relativity (GR) in the strong-field, highly dynamic, and relativistic regimes. Within the parameterized post-Newtonian (PN) formalisms, waveform generation tests have conventionally been limited to constraining inspiral coefficients up to the 3.5PN order. Leveraging the recent theoretical breakthrough that extended the analytical compact binary phasing to the 4.5PN order, we present the first observational constraints on these higher-order effects. Our analysis utilizes two exceptional events detected by the LIGO-Virgo-KAGRA (LVK) network: GW250114\_082203, which boasts the highest signal-to-noise ratio (SNR) recorded to date, and GW230627\_015337, which features a uniquely prolonged inspiral phase and the highest inspiral phase SNR to date. By performing Bayesian inference on the dimensionless deviation parameters ($\delta\phi_i$) associated with the 4PN and 4.5PN coefficients, we find that our results are fully consistent with the predictions of GR. While the current 90\% credible intervals for the four deviation parameters are of order $\mathcal{O}(1) \text{-} \mathcal{O}(10)$, the general relativistic null values ($\delta\hat{\phi}_a= 0$) are entirely encapsulated within the bounds. This investigation establishes the first empirical baseline for 4PN and 4.5PN inspiral tests of GR, paving the way for high-precision null tests of GR with current and next-generation GW detectors.

\end{abstract}

\maketitle
%\tableofcontents
\section{\label{sec:level1}Introduction}

General relativity (GR), Einstein’s theory of gravity, remains the most accurate and empirically robust framework for describing gravitational phenomena. It has passed stringent tests across a wide range of regimes, including precision measurements in the Solar System \cite{Will:2014kxa}, observations of binary pulsars \cite{Kramer:2021jcw, Freire:2024adf}, cosmological surveys \cite{Psaltis:2008bb, Clifton:2011jh, Ferreira:2019xrr}, and direct detections of gravitational waves (GWs) \cite{LIGOScientific:2016lio, LIGOScientific:2019fpa, LIGOScientific:2021sio, LIGOScientific:2020tif, LIGOScientific:2025wao, LIGOScientific:2026qni}. GWs, one of the key predictions of GR, are time-varying perturbations of spacetime that propagate at the speed of light. Observations of coalescing binary black holes (BBHs) \cite{LIGOScientific:2025hdt, LIGOScientific:2025slb}, as well as binary neutron stars and mixed binaries, have established a unique laboratory for testing GR in the strong-field, dynamical, and highly relativistic regime \cite{LIGOScientific:2025wao}. The event GW150914\_082203, first directly detected by LIGO-Virgo detectors in 2015 \cite{LIGOScientific:2016aoc}, provided the first opportunity to test GR in the extreme gravity regime and constrain its predictions \cite{LIGOScientific:2016lio}, the violations of its basic symmetry \cite{Zhu:2023rrx, Wang:2025fhw, Bian:2025ifp}, as well as those of alternative theories of gravity \cite{Yunes:2016jcc}. Over the past decade, the LIGO-Virgo-KAGRA (LVK) detector network has identified about 390 confident GW events \cite{LIGOScientific:2025slb, LIGOScientific:2026wfs}, including several exceptional sources \cite{LIGOScientific:2017ycc, LIGOScientific:2020stg, LIGOScientific:2020zkf, LIGOScientific:2020iuh, LIGOScientific:2024elc, LIGOScientific:2025cmm, LIGOScientific:2025rid, LIGOScientific:2025rsn, LIGOScientific:2025brd} whose unique properties have enabled particularly stringent tests of GR.

Tests of GR with GW observations are generally divided into generic tests and theory-specific tests. Generic tests, often referred to as null tests of GR, treat Einstein’s theory as the null hypothesis and search for possible deviations. Among the generic tests based on GW data, generation tests constitute an important class, as they probe whether the generation of the signal at the source is consistent with the predictions of GR. Two widely used frameworks for generation tests are the parameterized post-Newtonian (PPN) approach \cite{Arun:2006yw} and the parametrized post-Einsteinian  framework \cite{Yunes:2009ke}. In this paper, we focus exclusively on the PPN test. 

The PPN framework was previously explored in ref.~\cite{Arun:2006yw} and has since been extended and refined by refs.~\cite{Meidam:2017dgf, Agathos:2013upa, Roy:2025gzv, Mishra:2010tp, Mehta:2022pcn}. In this approach, the inspiral waveform is written as a post-Newtonian (PN) expansion, with phase coefficients ordered in powers of $v/c$ and derived perturbatively from the two-body dynamics and gravitational radiation \cite{Arun:2006yw}. For non-spinning binaries on quasi-circular orbits, these coefficients were known up to 3.5PN order until the recent works of Blanchet et al. \cite{Blanchet:2023bwj, Blanchet:2023sbv}, which extended the calculation to 4.5PN order, including certain nonlinear effects \cite{Trestini:2023wwg, Blanchet:2022vsm, Trestini:2022tot, Blanchet:2023bwj, Blanchet:2023sbv}.

Higher-order PN coefficients directly probe the nonlinear structure of GR and therefore provide a potential avenue for high-precision tests. However, the cumulative contribution of the 4PN and 4.5PN terms to the inspiral phase is small, of order one-tenth of a cycle, or less than 1 radian in the phase \cite{Blanchet:2023bwj}. Given the limited sensitivity of LVK detectors, which leads to relatively low signal-to-noise ratios (SNRs) for current detections, meaningful constraints on deviation parameters at these orders are not yet achievable. Recently, the authors of ref.~\cite{DuttaRoy:2024qbl} used the Fisher information matrix to forecast the constraining power of Advanced LIGO, Cosmic Explorer, Einstein Telescope, and LISA on newly introduced deviation parameters at 4PN and 4.5PN order. Their results indicate that current detectors provide only limited sensitivity to these coefficients, whereas future detectors may achieve substantially improved bounds. Since Fisher matrix approach forecasts assume stationary Gaussian noise and the high signal-to-noise ratio (SNR) limit, they are generally optimistic. Correspondingly, existing analyses of real GW data within the PPN framework have constrained inspiral coefficients only up to 3.5PN order \cite{LIGOScientific:2026fcf}.

Notably, the LVK detectors observed GW250114\_082203 (hereafter GW250114), the loudest GW signal to date, with a network SNR of 76 \cite{LIGOScientific:2025rid}. The exceptionally high SNR of GW250114 enables a more detailed characterization of the source and is particularly well suited for probing higher-order PN corrections during the inspiral phase \cite{Andres-Carcasona:2025bni}. In addition, GW230627\_015337 (hereafter GW230627), one of the events observed during the first part of the fourth observing run (O4a), has relatively lower component masses, $8.4_{-1.3}^{+1.6}\,M_\odot$ and $5.78_{-0.88}^{+0.96}\,M_\odot$, implying a long inspiral phase, and a comparatively high network SNR of $28.5_{-0.1}^{+0.1}$ \cite{LIGOScientific:2025slb, LIGOScientific:2026qni}. It is also the highest inspiral-phase SNR among all O4a events, yielding tighter constraints on several of the inspiral coefficients up to 3.5PN order \cite{LIGOScientific:2026fcf}. These two events are therefore exceptionally well suited for inspiral tests of GR. In this work, we perform Bayesian inference on deviation parameters associated with the 4PN and 4.5PN inspiral coefficients using data from both events. Our analysis provides meaningful constraints on these higher-order coefficients and serves as a reference for future efforts to test GR with higher-order inspiral terms using GW data. Looking ahead, the next generation of ground-based detectors such as Cosmic Explorer and the Einstein Telescope, as well as space-based GW detectors, will probe lower frequencies and achieve substantially improved sensitivity, enabling longer inspiral observations, higher SNRs, and consequently tighter constraints on higher-order inspiral coefficients within the PPN framework.

In this paper, we report tests of GR within the parametrized post-Newtonian framework at 4PN and 4.5PN order using data from the GW events GW250114 and GW230627. This paper is organized as follows. Section II introduces the parametrized waveform model at 4.5PN order. Section III outlines the Bayesian inference methodology employed to analyze the modified waveforms. Section IV presents the results, and Section V provides a summary and conclusions. 

\section{Parametrized waveform at the fourth post-Newtonian order}

Within the PN framework, an analytic expression for the inspiral GW phase can be derived in the slow-motion, weak-field regime, where the binary components are sufficiently separated and $v/c \ll 1$. For the dominant $(l=2, m=2)$ mode with aligned spins considered in this work, the frequency-domain waveform is obtained by applying the stationary phase approximation \cite{Damour:2000gg, Cutler:1994ys} to the Fourier transform of the time-domain signal. The inspiral phase, expanded up to 4.5PN order, is given by \cite{Blanchet:2023bwj, Blanchet:2023sbv}
\begin{eqnarray}
\Phi_{\mathrm{insp}}(f) &=& 2\pi f t_c - \phi_c + \frac{3}{128\eta v^5} \sum_{k=0}^{9} \biggl[ \phi_k v^k   \nonumber \\
&&~~ +\phi_{kl} v^k \ln v + \phi_{kl^2} v^k \ln^2 v + \cdots \biggr].
\end{eqnarray}
Here, $t_c$ and $\phi_c$ denote the coalescence time and phase, respectively. The PN expansion parameter is $v = (\pi M f)^{1/3}$, where $M=m_1+m_2$ is the total mass and $f$ is the GW frequency. We define the mass ratio as $q=m_1/m_2$, and the symmetric mass ratio as $\eta = m_1m_2/M^2 = q/(q+1)^2$, with $m_1 \geq m_2$. Throughout this paper, we use geometrized units with $G=c=1$. The leading-order term, corresponding to $k=0$, is the Newtonian (0PN) contribution; in our notation, any term proportional to $v^k$ is assigned to $\frac{k}{2}$PN order.

At 4PN and 4.5PN order \cite{Blanchet:2023bwj, Blanchet:2023sbv}, the GW phenomenology of compact binaries acquires several important nonlinear contributions. The familiar tail and tail-of-tail interactions are extended to quartic order, and the 4.5PN energy flux confirms the expected tail-of-tail-of-tail structure \cite{Marchand:2016vox}. At 4PN order, the nonlinear memory effect induces the tail-of-memory contribution to the $(\ell,m)=(2,0)$ mode of the GW signal \cite{Trestini:2022tot, Blanchet:2023bwj}. For the dominant $(2,2)$ mode, the required tail integrals at 4PN can be consistently interpreted within a renormalization-group framework, leading to a scale-dependent Bondi mass. In this picture, the logarithmic term in the 4PN binding energy arises as a renormalization-group effect \cite{Galley:2015kus}.

The newly computed terms at 4PN and 4.5PN introduce additional structure into the inspiral phasing. In particular, logarithmic contributions arise at both 4PN and 4.5PN order, along with a $\ln^2 v$ term at 4PN and a non-logarithmic term at 4.5PN. Moreover, the non-logarithmic 4PN term can be absorbed into a redefinition of the coalescence time $t_c$. To highlight the structure of these newly computed phasing terms, we rewrite the inspiral phase as follows \cite{DuttaRoy:2024qbl}:
\begin{eqnarray}
\Phi_{\mathrm{insp}}(f) &=& 2\pi f t_c - \phi_c + \frac{3}{128\,\eta\, v^{5}} \Big[\Phi_{3.5\mathrm{PN}}\nonumber\\
&&\quad + v^{8} \big( \phi_{8\ell} \ln v + \phi_{8\ell^2} \ln^2 v \big) \nonumber\\
&&\quad + v^{9} \big( \phi_{9} + \phi_{9\ell} \ln v \big)\Big],
\label{eq:insp2} 
\end{eqnarray}
where $\Phi_{3.5\mathrm{PN}}$ denotes the 3.5PN phasing normalized to the Newtonian leading-order term, and the remaining terms correspond to the new 4PN and 4.5PN coefficients. The explicit expressions for $\phi_{8\ell}$, $\phi_{8\ell^{2}}$, $\phi_{9}$, and $\phi_{9\ell}$, which can be found in \cite{Blanchet:2023bwj, Blanchet:2023sbv}, are listed as follows:
\begin{eqnarray}
\phi _{8\ell} &=& -\frac{2550713843998885153}{276808510218240} +\frac{90490}{189} \pi ^{2} \nonumber\\ 
&& +\frac{36812}{63} \gamma _{E} + \frac{1011020}{1323} \ln{2} +\frac{78975}{49} \ln{3}\nonumber\\
&& +\eta\left(\frac{680712846248317}{42247941120}-\frac{109295}{224}\pi^{2} \right.\nonumber\\
&&~~~~~~~ +\left.\frac{3911888}{1323}\gamma_{E}+\frac{9964112}{1323}\ln{2} \right.\nonumber\\ 
&&~~~~~~~ -\left.\frac{78975}{49}\ln{3} \right)\nonumber\\
&& + \eta^{2}\left(-\frac{7510073635}{3048192}+\frac{11275}{144}\pi^{2}\right) \nonumber\\ 
&&- \frac{1292395}{12096}\eta^{3}+\frac{5975}{96}\eta^{4},
\end{eqnarray}
\begin{eqnarray}
\phi _{8\ell^{2}} = \frac{18406}{63} +\frac{1955944}{1323} \eta,
\end{eqnarray}
\begin{eqnarray}
\phi _{9} &=& \pi\left[\frac{105344279473163}{18776862720} -\frac{640}{3} \pi ^{2} \right.\nonumber\\
&&~~\left.\quad-\frac{13696}{21} \gamma _{E}-\frac{13696}{21}\ln{4}\right.\nonumber\\ 
&&~~~~~\left.+\eta\left(-\frac{149291720735}{134120448}+\frac{2255}{6}\pi ^{2}\right)\right.\nonumber\\
&&~~\left.\quad+\frac{45293335}{127008}\eta ^{2}+\frac{10323755}{199584}\eta ^{3}\right],\nonumber\\
\end{eqnarray}
and
\begin{eqnarray}
\phi _{9\ell}  = -\frac{13696}{21}\pi.           
\end{eqnarray}

In parametrized tests of GR within the PPN framework, the GR waveform is modified by introducing phenomenological deviations at each PN order of the inspiral phase, in the form $\phi_a \rightarrow \phi_a^{\mathrm{GR}}(1+\delta\hat{\phi}_a)$. Here, $\delta\hat{\phi}_a \equiv \delta\hat{\phi}_a/\phi_a$ denotes the fractional deviation from the GR value. By construction, $\delta\hat{\phi}_a=0$ corresponds to GR; consequently, if the posterior distribution inferred from a compact-binary signal is consistent with zero, the signal is deemed statistically consistent with GR.  

When the inspiral phase is expanded up to 4.5PN order, each coefficient may be labeled as $\phi_a$, where $a=\{k,k\ell,k\ell^2\}$ denotes, respectively, the non-logarithmic, logarithmic, and squared-logarithmic contributions. The PN coefficients $(\phi_k,\phi_{k\ell},\phi_{k\ell^2})$ are then deformed according to
\begin{eqnarray}
\phi_{k} &\rightarrow& \phi_{k}\left(1+\delta\hat{\phi}_{k}\right), \nonumber\\
\phi_{k\ell} &\rightarrow& \phi_{k\ell}\left(1+\delta\hat{\phi}_{k\ell}\right), \nonumber\\
\phi_{k\ell^{2}} &\rightarrow& \phi_{k\ell^{2}}\left(1+\delta\hat{\phi}_{k\ell^{2}}\right),
\end{eqnarray}
where the corresponding fractional deviation parameters vanish in GR. In the spirit of parametrized tests, we further introduce four null parameters at 4PN and 4.5PN order: at 4PN, the logarithmic and squared-logarithmic parameters $\delta\hat{\phi}_{8\ell}$ and $\delta\hat{\phi}_{8\ell^2}$, and at 4.5PN, the non-logarithmic and logarithmic parameters $\delta\hat{\phi}_{9}$ and $\delta\hat{\phi}_{9\ell}$. The resulting parameterized phasing at 4PN and 4.5PN is therefore specified by the set $\{\delta\hat{\phi}_{8\ell},\delta\hat{\phi}_{8\ell^2},\delta\hat{\phi}_{9},\delta\hat{\phi}_{9\ell}\}$.

\section{Bayesian inference on the modified waveforms}

In this section, we describe the Bayesian inference for constraining the deviation parameters at 4PN and 4.5PN order in the inspiral phase of GWs within the PPN framework. In particular, we focus on the four deviation parameters
$\{\delta\hat{\phi}_{8\ell},\; \delta\hat{\phi}_{8\ell^2},\; \delta\hat{\phi}_{9},\; \delta\hat{\phi}_{9\ell}\}$,
which parametrize departures from the GR inspiral phasing at 4PN and 4.5PN order.

Bayesian inference is widely used in GW astronomy for both parameter estimation and model selection. In our analysis, given a GW data stream $d_i$, we compare the observed data with the waveform model $h(\vec{\theta})$ under a parametrized waveform hypothesis $H$ that includes possible modifications. According to Bayes’ theorem, the posterior distribution of the model parameters $\vec{\theta}$ is given by
\begin{equation}
P(\vec{\theta}\mid d,H)=\frac{P(d\mid \vec{\theta},H)\,P(\vec{\theta}\mid H)}{P(d\mid H)} ,
\end{equation}
where $P(\vec{\theta}\mid H)$ is the prior distribution, $P(d\mid \vec{\theta},H)$ is the likelihood, and $P(d\mid H)$ is the Bayesian evidence. Assuming stationary Gaussian detector noise, the likelihood can be written in matched-filtering form as
\begin{equation}
P(d\mid \vec{\theta},H)\propto \prod_{i=1}^{n}\exp\!\left[-\frac{1}{2}\left\langle d_i-h(\vec{\theta})\,\middle|\, d_i-h(\vec{\theta})\right\rangle\right],
\end{equation}
where the noise-weighted inner product is defined as
\begin{equation}
\left\langle A\mid B\right\rangle
=4\,\mathrm{Re}\!\left[\int_{0}^{\infty}\frac{A(f)B^{\ast}(f)}{S_n(f)}\,df\right].
\end{equation}
Here, $^\ast$ denotes complex conjugation and $S_n(f)$ is the detector noise power spectral density (PSD). In our analysis, we use the PSDs provided in the LVK's parameter estimation release \cite{LIGOScientific:2025slb, LIGOScientific:2025rid}, which are estimated using the Welch averaging \cite{Cornish:2014kda, Littenberg:2014oda}.

To test the PN phase coefficients, we employ a modified waveform obtained from the GR template $\tilde{h}_A^{\mathrm{GR}}(f)$ by introducing a frequency-domain phase correction. Specifically, the modified polarization modes are written as
\begin{equation}
\tilde{h}_A(f)=\tilde{h}_A^{\mathrm{GR}}(f)\,e^{i\,\delta\Psi(f)}, \label{waveform}
\end{equation}
where $A=\{+,\times\}$ denotes the plus and cross polarizations. For the 4PN and 4.5PN modifications considered in this work, the dephasing term is given by
\begin{eqnarray}
\delta\Psi(f)&=&\frac{3}{128\,\eta\,v^{5}}
\Big[
v^{8}\big(\phi_{8\ell}\delta\hat{\phi}_{8\ell}+\phi_{8\ell^2}\delta\hat{\phi}_{8\ell^2}\ln v\big) \nonumber\\
&&\;\;\;\;\;\;\;\;\;\;\;\;\;\;\;+v^{9}\big(\phi_{9}\delta\hat{\phi}_{9}+\phi_{9\ell}\delta\hat{\phi}_{9\ell}\ln v\big)
\Big],
\end{eqnarray}
with $v=(\pi Mf)^{1/3}$, $M$ the total mass, and $\eta$ the symmetric mass ratio. The coefficients $\phi_{8\ell}$, $\phi_{8\ell^2}$, $\phi_{9}$, and $\phi_{9\ell}$ are the GR values of the corresponding 4PN and 4.5PN phasing terms. The circular polarization modes $\tilde{h}_R$ and $\tilde{h}_L$ are related to $\tilde{h}_+$ and $\tilde{h}_\times$ through the usual linear combinations. Eq.~(\ref{waveform}) represents the modified waveform used for comparison with the GW data. The tests are carried out using the open data of the binary black hole merger events GW230627 and GW250114.

We perform Bayesian parameter estimation using the open-source package \textsc{Bilby} \cite{Ashton:2018jfp, Romero-Shaw:2020owr} and the sampler \textsc{nessai} \cite{williams2024nessai, Williams:2021qyt, Williams:2023ppp}. To reduce modeling systematics, we consider two GR-based waveform templates: \textsc{SEOBNRv5HM\_ROM} \cite{Pompili:2023tna}, which is restricted to aligned-spin configurations, and \textsc{IMRPhenomXPHM} \cite{Pratten:2020ceb, Colleoni:2024knd}, which includes precession effects. The priors on the standard binary parameters (masses, spins, distance, inclination, etc.) are chosen consistently with the LVK posterior releases for each event, while each deviation parameter $\delta\hat{\phi}_a$ is assigned a uniform prior over a sufficiently broad interval (e.g., $[-80,80]$) to avoid prior-induced bias. The priors adopted in all the analyses are listed in Tables \ref{mass_distance}, \ref{spin_parameters}, and \ref{deviation_parameters} in Appendix A.

For the parameter estimation, we impose a minimum frequency of $20\,\mathrm{Hz}$ and a maximum frequency cutoff $f_{\rm c}$ corresponding to the transition from the inspiral to the intermediate regime, defined by $f_{\rm c}=0.018 c^3/(G M)$ \cite{LIGOScientific:2019fpa}. This choice ensures that the analysis remains primarily sensitive to the inspiral portion of the signal, where the PN approximation is expected to be most reliable. We also performed analyses with a choice of the cutoff  $f_{\rm max}=0.021\,c^3/(GM)$, and the resulting posterior distributions of the four derivative parameters are summarized in Appendix~B. Here we would like to mention that, following the approach adopted in most related studies in the literature \cite{LIGOScientific:2016lio, LIGOScientific:2019fpa, LIGOScientific:2021sio, LIGOScientific:2020tif, LIGOScientific:2025wao, LIGOScientific:2026qni}, we constrain one deviation parameter at a time, while fixing all other PN deviation parameters to zero. In other words, only a single $\delta\hat{\phi}_a$ is allowed to vary jointly with the standard GR parameters in each analysis.

\section{Results and discussion}

We present constraints on the four deviation parameters $\delta\hat{\phi}_{8\ell}$, $\delta\hat{\phi}_{8\ell^{2}}$, $\delta\hat{\phi}_{9}$, and $\delta\hat{\phi}_{9\ell}$ inferred from GW250114 and GW230627 within the PPN framework. For each event, we analyze the data with the SEOBNRv5HM\_ROM and IMRPhenomXPHM waveform models. In each analysis, only one deviation parameter is allowed to vary, while the others are fixed to their GR values. The resulting medians, 90\% credible intervals, and GR quantiles $Q_{\mathrm{GR}} = P(\delta\hat{\phi}_a < 0)$ are summarized in Table~\ref{tab:results}. $P(\delta\hat{\phi}_a < 0)$ is the probability that a random draw from the posterior distribution of $\delta\hat{\phi}_a$ takes a negative value. Values of $Q_{\mathrm{GR}}$ close to 50\% indicate consistency with the null hypothesis.

\begin{table*}
\caption{Median values, 90\% credible intervals, and GR quantiles $Q_{\mathrm{GR}}$ for the deviation parameters $\delta\hat{\phi}_a$ inferred from GW230627 and GW250114 using the SEOBNRv5HM\_ROM and IMRPhenomXPHM waveform models. The GR quantile is defined as $Q_{\mathrm{GR}} = P (\delta\hat{\phi}_a < 0)$, where $\delta\hat{\phi}_a < 0$ corresponds to the GR value.}.
\label{tab:results}
\begin{ruledtabular}
\begin{tabular}{lcccccccc}   
\multicolumn{1}{c}{} &
\multicolumn{4}{c}{GW230627} &
\multicolumn{4}{c}{GW250114} \\
\cline{2-5} \cline{6-9} &
\multicolumn{2}{c}{SEOBNRv5HM\_ROM} &
\multicolumn{2}{c}{IMRPhenomXPHM} &
\multicolumn{2}{c}{SEOBNRv5HM\_ROM} &
\multicolumn{2}{c}{IMRPhenomXPHM} \\
Parameters & \multicolumn{1}{c}{$\delta\hat{\phi}_a$} & $Q_{\rm GR}$ & \multicolumn{1}{c}{$\delta\hat{\phi}_a$} & $Q_{\rm GR}$ & \multicolumn{1}{c}{$\delta\hat{\phi}_a$} & $Q_{\rm GR}$ & \multicolumn{1}{c}{$\delta\hat{\phi}_a$} & $Q_{\rm GR}$ \\
\midrule
$\delta\hat{\phi}_{8\ell}$& $-0.586^{+9.378}_{-6.781}$& 56.80\% & $-0.225^{+3.802}_{-3.802}$& 53.80\% & $5.58^{+7.961}_{-7.268}$& 9.20\%  & $6.067^{+7.083}_{-6.434}$& 5.80\% \\
$\delta\hat{\phi}_{8\ell^{2}}$& $0.447^{+2.999}_{-3.442}$& 40.90\% & $0.605^{+2.743}_{-3.069}$& 36.00\% & $-3.969^{+6.6}_{-8.02}$& 85.00\% & $-3.836^{+5.143}_{-5.547}$& 89.50\% \\
$\delta\hat{\phi}_{9}$    & $-0.994^{+5.168}_{-4.556}$& 63.60\% & $0.277^{+4.828}_{-4.623}$& 46.00\% & $2.129^{+3.029}_{-6.32}$& 24.40\% & $2.644^{+2.557}_{-5.29}$& 17.00\% \\
$\delta\hat{\phi}_{9\ell}$& $6.455^{+13.461}_{-15.915}$& 23.80\% & $6.352^{+13.317}_{-15.598}$& 24.20\% & $-5.453^{+7.881}_{-4.82}$& 86.00\% & $-5.331^{+7.324}_{-4.843}$& 87.80\% 
\end{tabular}
\end{ruledtabular}
\end{table*}

For both events, the posterior distributions of all four deviation parameters are consistent with GR. In particular, for GW230627, $\delta\hat{\phi}_{8\ell^{2}}$ is the most tightly constrained parameter, with medians $0.605$ and $0.447$ and 90\% credible intervals $[-2.464, 3.348]$ and $[-2.995, 3.446]$ for IMRPhenomXPHM and SEOBNRv5HM\_ROM, respectively. For GW250114, $\delta\hat{\phi}_{9}$ is the most tightly constrained, with medians $0.277$ and $-0.994$ and 90\% credible intervals $[-4.346, 5.105]$ and $[-5.550, 4.174]$. The two waveform families yield consistent results, with IMRPhenomXPHM giving marginally tighter bounds than SEOBNRv5HM\_ROM.

Overall, GW230627 provides stronger constraints than GW250114 on $\delta\hat{\phi}_{8\ell}$, $\delta\hat{\phi}_{8\ell^{2}}$, and $\delta\hat{\phi}_{9}$. This is reflected in medians closer to zero, narrower credible intervals, and $Q_{\mathrm{GR}}$ values closer to 50\%. The only exception is $\delta\hat{\phi}_{9\ell}$, for which GW230627 yields weaker constraints than GW250114. This trend is likely driven by the longer inspiral duration of GW230627, despite the higher SNR of GW250114.

\begin{figure*}
\includegraphics[width=14cm]{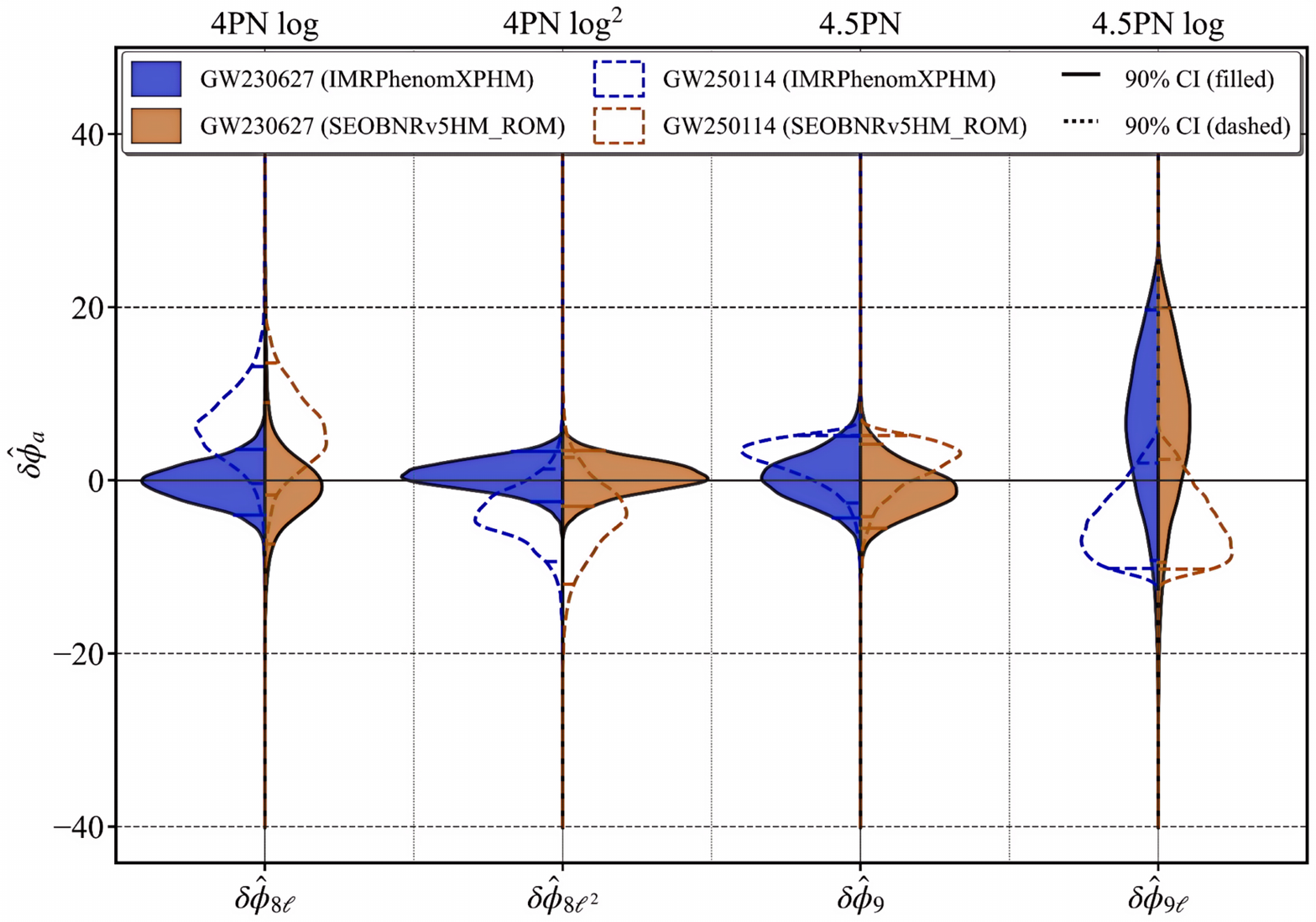}
\caption{Violin plots of the posterior distributions for the four deviation parameters. The left half of each violin corresponds to constraints obtained with the IMRPhenomXPHM template for GW230627 and GW250114, while the right half corresponds to the SEOBNRv5HM\_ROM template. Filled colored areas represent results for GW230627; unfilled areas with dashed lines represent results for GW250114.
}
\label{fig:phi_bounds}
\end{figure*}

Figure~\ref{fig:phi_bounds} shows the posterior distributions of the four deviation parameters. The trends discussed above are more clearly illustrated in this figure. For GW230627, the posteriors of $\delta\hat{\phi}_{8\ell}$, $\delta\hat{\phi}_{8\ell^{2}}$, and $\delta\hat{\phi}_{9}$ are more concentrated and more symmetrically distributed around $\delta\hat{\phi}_a= 0$ than those for GW250114, indicating tighter constraints and better agreement with GR. The only exception is $\delta\hat{\phi}_{9\ell}$, for which GW230627 yields weaker constraints than GW250114. Nevertheless, $\delta\hat{\phi}_{9\ell}=0$ lies within the 90\% credible intervals for both events.

\begin{figure*}
\includegraphics[width=14cm]{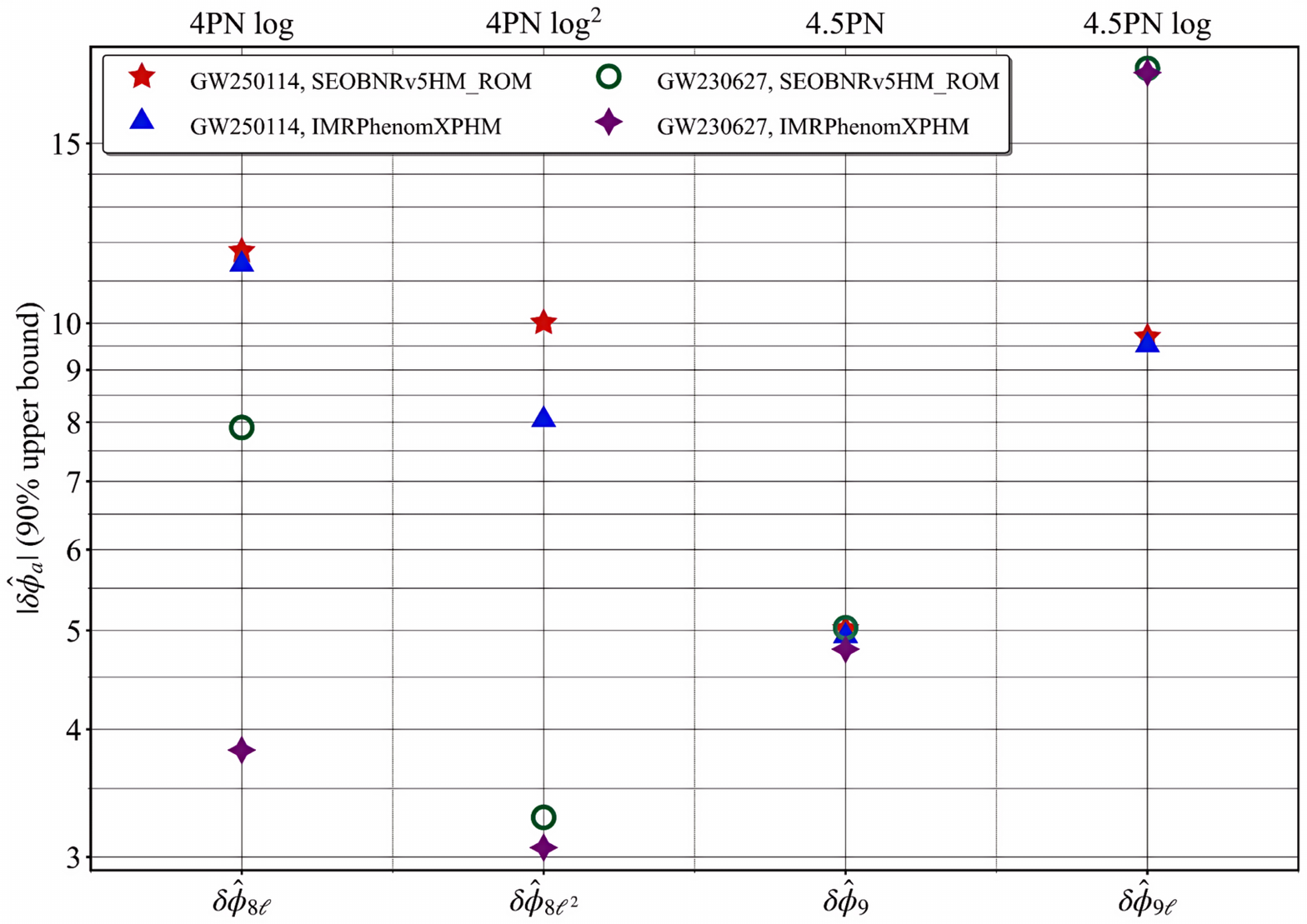}
\caption{90\% upper bounds on the magnitudes of the deviation coefficients $\{\delta\hat{\phi}_{8\ell}, \delta\hat{\phi}_{8\ell^2}, \delta\hat{\phi}_{9}, \delta\hat{\phi}_{9\ell}\}$ inferred from GW230627 and GW250114 using the SEOBNRv5HM\_ROM and IMRPhenomXPHM waveform models. The purple stars denote bounds from GW250114 with SEOBNRv5HM\_ROM; the green triangles denote bounds from GW250114 with IMRPhenomXPHM; the blue open circles denote bounds from GW230627 with SEOBNRv5HM\_ROM; and the red diamonds denote bounds from GW230627 with IMRPhenomXPHM.}
\label{fig:phi_bounds2}
\end{figure*}

Figure~\ref{fig:phi_bounds2} presents the 90\% upper bounds of the posterior distributions for the four deviation parameters. These upper bounds are mostly in the range $3-10$, with two exceptions: the bound on $\delta\hat{\phi}_{8\ell}$ from GW250114 and the bound on $\delta\hat{\phi}_{9\ell}$ from GW230627. For $\delta\hat{\phi}_{8\ell}$, $\delta\hat{\phi}_{8\ell^{2}}$, and $\delta\hat{\phi}_{9}$, the upper bounds from GW230627 are lower than those from GW250114. The tightest bounds for these three parameters are obtained from GW230627 with the IMRPhenomXPHM waveform model, consistent with the results above. Among all parameters, $\delta\hat{\phi}_{9}$ shows the best consistency across events and waveform models, with all upper bounds close to 5. This may be because $\delta\hat{\phi}_{9}$ contains no logarithmic term, unlike the other parameters, which reduces systematic effects. The fact that most upper bounds fall within $1-10$ likely reflects the small cumulative phase contribution of the 4PN and 4.5PN coefficients (less than one cycle \cite{Blanchet:2023bwj}), which limits how tightly they can be constrained with the present method. For $\delta\hat{\phi}_{9\ell}$, the upper bounds from GW230627 are about 17.6 for both waveform models, larger than those from GW250114 (around 9.6).

In summary, all constraints are consistent with GR. GW230627 generally provides tighter bounds than GW250114, especially for $\delta\hat{\phi}_{8\ell}$, $\delta\hat{\phi}_{8\ell^{2}}$, and $\delta\hat{\phi}_{9}$, while $\delta\hat{\phi}_{9\ell}$ remains comparatively weakly constrained.

\section{Conclusion and Outlook}

We have analyzed two GW events, GW250114 and GW230627, to constrain the four deviation parameters associated with the 4PN and 4.5PN inspiral-phase coefficients in the PPN expansion of the inspiral waveform. The analysis was performed using modified versions of the SEOBNRv5HM\_ROM and IMRPhenomXPHM waveform models, in which additional phase corrections were introduced to describe possible deviations from GR. Independent constraints were placed on $\delta\hat{\phi}_{8\ell}$, $\delta\hat{\phi}_{8\ell^{2}}$, $\delta\hat{\phi}_{9}$, and $\delta\hat{\phi}_{9\ell}$, while the remaining deviation parameters were fixed to their GR values. We restricted the analysis to the inspiral regime, from $20\,\mathrm{Hz}$ to $0.018\,c^3/(GM)$, and compared the data with waveforms including the corresponding 4PN and 4.5PN phase corrections.

Our results show that all inferred posteriors of the four derivation parameters are consistent with GR, with $\delta\hat{\phi}_a= 0$ lying within the 90\% credible interval in every case. The two waveform models yield mutually consistent constraints, with only minor differences. Among the four parameters, $\delta\hat{\phi}_{9}$ is the most consistently constrained across both events and both waveform models, with 90\% upper bounds close to 5. The 90\% upper bounds for most parameters lie in the range $3$--$10$, while $\delta\hat{\phi}_{9\ell}$ remains comparatively weakly constrained for GW230627.

Overall, GW230627 provides tighter constraints than GW250114 for $\delta\hat{\phi}_{8\ell}$, $\delta\hat{\phi}_{8\ell^{2}}$, and $\delta\hat{\phi}_{9}$, as reflected in medians closer to zero, narrower credible intervals, $Q_{\mathrm{GR}}$ values nearer to 50\%, and more concentrated posterior distributions. This suggests that the longer inspiral duration of GW230627 makes it particularly valuable for testing deviations from GR in the inspiral phase.

In future work, we will extend this analysis to additional high-SNR events with long inspiral signals to obtain more comprehensive Bayesian constraints on the 4PN and 4.5PN deviation parameters. We will also forecast the expected bounds from next-generation ground-based detectors such as CE and ET, as well as space-based detectors such as LISA.

\begin{acknowledgments}

This work is supported in part by the National Natural Science Foundation of China under Grants No. 12275238 and 11675143, the National Key Research and Development Program under Grant No. 2020YFC2201503, and the Zhejiang Provincial Natural Science Foundation of China under Grants No. LR21A050001 and No. LY20A050002, and the Fundamental Research Funds for the Provincial Universities of Zhejiang in China under Grant No. RF-A2019015. Wen Zhao is supported by the National Natural Science Foundation of China under Grant No. 12325301. Xin Zhang is supported by the National Natural Science Foundation of China (Grants Nos. 12473001, 12575049, and 12533001), the National SKA Program of China (Grants Nos. 2022SKA0110200 and 2022SKA0110203), the China Manned Space Program (Grant No. CMS-CSST-2025-A02), and the 111 Project (Grant No. B16009). 

This research has made use of data or software obtained from the Gravitational Wave Open Science Center (gwosc.org), a service of the LIGO Scientific Collaboration, the Virgo Collaboration, and KAGRA. This material is based upon work supported by NSF's LIGO Laboratory which is a major facility fully funded by the National Science Foundation, as well as the Science and Technology Facilities Council (STFC) of the United Kingdom, the Max-Planck-Society (MPS), and the State of Niedersachsen/Germany for support of the construction of Advanced LIGO and construction and operation of the GEO600 detector. Additional support for Advanced LIGO was provided by the Australian Research Council. Virgo is funded, through the European Gravitational Observatory (EGO), by the French Centre National de Recherche Scientifique (CNRS), the Italian Istituto Nazionale di Fisica Nucleare (INFN) and the Dutch Nikhef, with contributions by institutions from Belgium, Germany, Greece, Hungary, Ireland, Japan, Monaco, Poland, Portugal, Spain. KAGRA is supported by the Ministry of Education, Culture, Sports, Science and Technology (MEXT), Japan Society for the Promotion of Science (JSPS) in Japan; National Research Foundation (NRF) and Ministry of Science and ICT (MSIT) in Korea; Academia Sinica (AS) and National Science and Technology Council (NSTC) in Taiwan.

The data analyses and results visualization in this work made use of \texttt{BILBY} \cite{Ashton:2018jfp, Romero-Shaw:2020owr}, \texttt{nessai} \cite{williams2024nessai, Williams:2021qyt, Williams:2023ppp}, \texttt{LALSuite} \cite{lalsuite}, \texttt{Numpy} \cite{Harris:2020xlr, vanderWalt:2011bqk}, \texttt{Scipy} \cite{Virtanen:2019joe}, and \texttt{matplotlib} \cite{Hunter:2007ouj}.

\end{acknowledgments}

\appendix
\section{Priors used in the analyses}

The priors used in the analyses of the two GW events, GW230627 and GW250114, are summarized in the tables below, i.e., Tables \ref{mass_distance}, \ref{spin_parameters}, and \ref{deviation_parameters}. The names of the priors refer to the names in the Bilby package \cite{Ashton:2018jfp, Romero-Shaw:2020owr}. The priors of the chirp masses ${\cal M}_{\rm c}$, constraints on the component masses $m_1$ and $m_2$ in the detector frame,  the luminosity distances $d_L$, and the mass ratio $q=m_1/m_2$ are summarized in Table \ref{mass_distance} for GW events, GW230627 and GW250114, respectively. For each event, we employed two waveform models: IMRPhenomXPHM and SEOBNRv5HM\_ROM. The corresponding prior distributions of the spin parameters for these two waveform templates in the analysis are listed in Table \ref{spin_parameters}. For parameters associated with SEOBNRv5HM\_ROM in Table \ref{spin_parameters}, $\chi_1$ and $\chi_2$ are the aligned spin magnitudes of the two components, while for IMRPhenomXPHM, $a_1$ and $a_2$ represent the dimensionless spin magnitudes of the two components, $\phi_{12}$ is the difference between the azimuthal angles of the individual spin vector projections on to the orbital plane, and $\phi_{jl}$ represents the difference between the total and orbital angular momentum azimuthal angles. For both templates, $ \mathrm{ra} $ is the right ascension, $\psi$ the polarization angle of the source, and $\phi_{\rm ref}$ denotes the binary phase at the reference frequency. 

The priors for the four deviation parameters are given in Table \ref{deviation_parameters}. Unless otherwise specified, each deviation parameter is assigned a uniform prior over the interval $[-80,80]$. The only exception is $\delta\hat{\phi}_{9\ell}$ when analyzing GW250114 with the IMRPhenomXPHM waveform model: for this case, the prior range is narrowed to $[-50, 50]$. This narrowing is necessary because a prior range of $[-80, 80]$ leads to multi-modal posteriors, which are not physically realistic. Consequently, we reduced the prior range to avoid such behavior.

\begin{table}
\caption{The priors on the chirp mass and luminosity distances for two events in all analyses.}
\label{mass_distance}
\begin{ruledtabular}
\begin{tabular}{cll}
\textbf{Events} & \textbf{parameters} & \textbf{priors} \\
\hline
\multirow{4}{*}{GW230627}
& $m_1$ & constraint(1, 1000) \\
&$ m_2$ & constraint(1, 1000) \\
& $\mathcal{M}_c $& Uniform(6.39, 6.44) \\
%& $a_1$ & Uniform(0, 0.99) \\
%&$ a_2$ & Uniform(0, 0.99) \\
%& $\phi_{12}$ & Uniform(0, $2\pi$) \\
%&$ \phi_{1l} $& Uniform(0, $2\pi$) \\
&$ d_L$ & UniformSourceFrame(50, 471.6) \\
& $q$ & Uniform(0.05,1) \\
%&$ \mathrm{ra} $& Uniform(0, $2\pi$) \\
%&$ \psi $& Uniform(0, $\pi$) \\
%& $\phi_{\rm ref} $& Uniform(0, $2\pi$) \\
\hline   
\multirow{4}{*}{GW250114}
& $m_1$ & constraint(10, 80) \\
& $m_2$ & constraint(10, 80) \\
& $\mathcal{M}_c$ & Uniform(29.9, 32.1) \\
%& $\chi_1$ & AlignedSpin(0, 0.99) \\
%& $\chi_2$ & AlignedSpin(0, 0.99) \\
& $d_L$ & UniformSourceFrame(10, 10000) \\
& $q$ & Uniform(0.05,1) \\
%&$ \mathrm{ra}$ & Uniform(0, $2\pi$) \\
%&$ \psi $& Uniform(0, $\pi$) \\
%&$ \phi_{\rm ref}$ & Uniform(0, $2\pi$) 
\end{tabular}
\end{ruledtabular}
\end{table}

\begin{table}
\caption{The priors on the spin parameters used for the different waveform approximants in each analysis with both events GW230627 and GW250114. The alignedSpin prior gives the prior distribution of the aligned spin component based on the generic spin priors.}
\label{spin_parameters}
\begin{ruledtabular}
\begin{tabular}{cll}
\textbf{Waveforms} & \textbf{parameters} & \textbf{priors} \\
\hline
\multirow{7}{*}{IMRPhenomXPHM}
%& m_1 & constraint(1, 1000) \\
%& m_2 & constraint(1, 1000) \\
%& \mathcal{M}_c & Uniform(6.38537095164235, 6.4428834107576165) \\
& $a_1$ & Uniform(0, 0.99) \\
&$ a_2$ & Uniform(0, 0.99) \\
& $\phi_{12}$ & Uniform(0, $2\pi$) \\
&$ \phi_{jl} $& Uniform(0, $2\pi$) \\
%& d_L & bilby.gw.prior.UniformSourceFrame(50, 471.599) \\
&$ \mathrm{ra} $& Uniform(0, $2\pi$) \\
&$ \psi $& Uniform(0, $\pi$) \\
& $\phi_{\rm ref} $& Uniform(0, $2\pi$) \\
\hline   
\multirow{5}{*}{SEOBNRv5HM\_ROM}
%& m_1 & constraint(1, 1000) \\
%& m_2 & constraint(1, 1000) \\
%& \mathcal{M}_c & Uniform(6.38537095164235, 6.4428834107576165) \\
& $\chi_1$ & alignedSpin(0, 0.99) \\
& $\chi_2$ & alignedSpin(0, 0.99) \\
%& d_L & bilby.gw.prior.UniformSourceFrame(50, 471.599) \\
&$ \mathrm{ra}$ & Uniform(0, $2\pi$) \\
&$ \psi $& Uniform(0, $\pi$) \\
&$ \phi_{\rm ref}$ & Uniform(0, $2\pi$) 
\end{tabular}
\end{ruledtabular}
\end{table}

\begin{table}
\caption{Prior distributions for the four deviation parameters for most of the runs and waveforms except for the analysis of GW250114 with the \textsc{IMRPhenomXPHM} waveform model; the prior for $\delta\hat{\phi}_{9\ell}$ is set to Uniform($-50.0$, $50.0$ for this case.}
\label{deviation_parameters}
\begin{ruledtabular}
\begin{tabular}{ll}
\textbf{Deviation parameters} & \textbf{priors} \\
\midrule
$\delta\hat{\phi}_{8\ell}$& Uniform($-80.0$, $80.0$) \\
$\delta\hat{\phi}_{8\ell^{2}}$& Uniform($-80.0$, $80.0$) \\
$\delta\hat{\phi}_{9}$& Uniform($-80.0$, $80.0$) \\
$\delta\hat{\phi}_{9\ell}$& Uniform($-80.0$, $80.0$) \\
\end{tabular}
\end{ruledtabular}
\end{table}

\section{Comparision between different cutoff frequencies}

In the main text, we present the results obtained by restricting the inspiral analysis to the frequency range from $20\,\mathrm{Hz}$ to $f_c = 0.018\,c^3/(GM)$. Since the choice of the inspiral cutoff frequency may affect the inferred constraints on the deviation parameters, we also investigate the robustness of our results by repeating the analysis with a slightly higher cutoff frequency, $f_{\max} = 0.021\,c^3/(GM)$.

\begin{figure*}
\includegraphics[width=16cm]{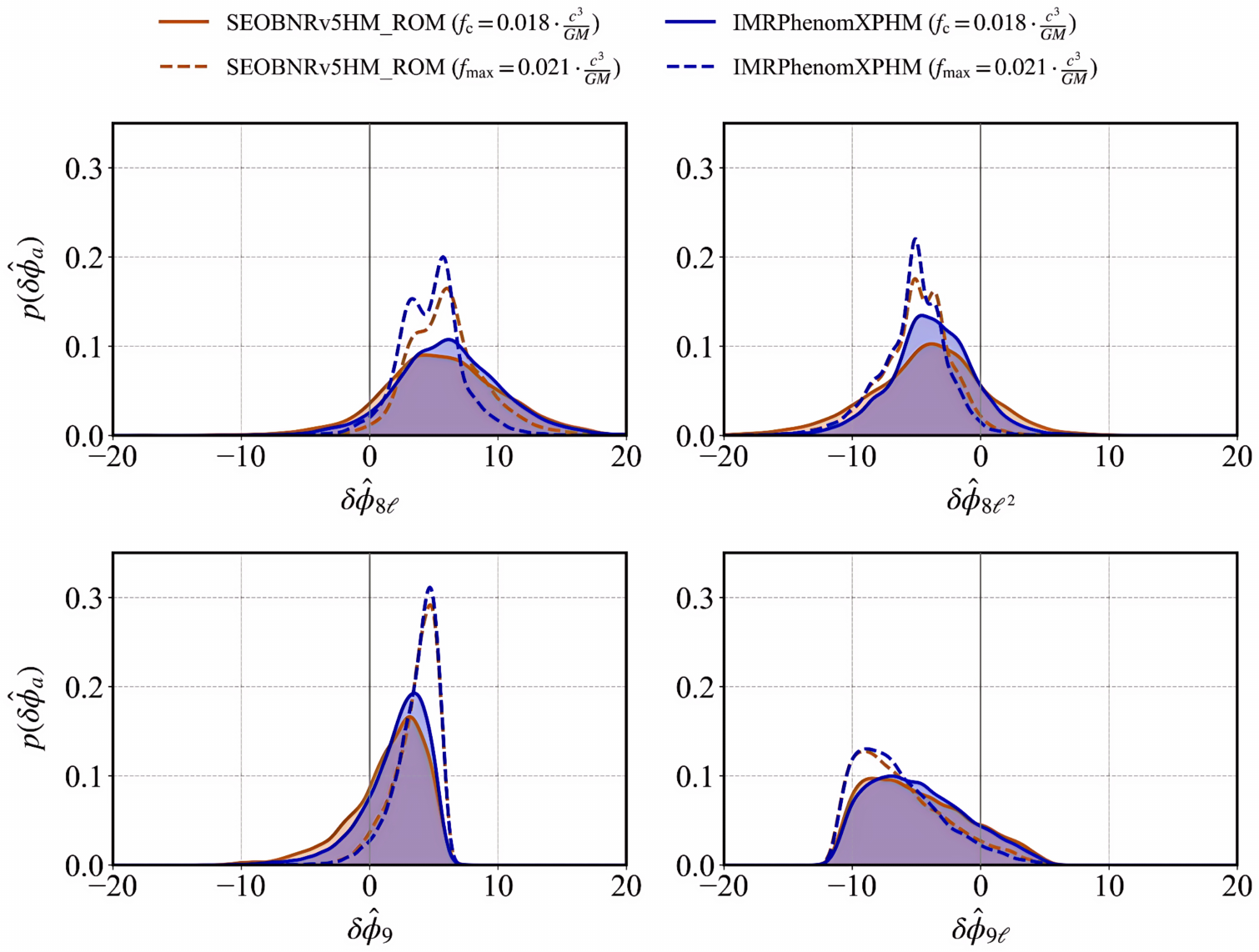}
\caption{Posterior distributions of the deviation parameters $\delta\hat{\phi}_a$ for GW250114 obtained using two differenet cutoff frequencies $f_c = 0.018\,c^3/(GM)$  and $f_{\rm max} = 0.021\,c^3/(GM)$.}
\label{fig:phi_bounds3}
\end{figure*}

\begin{figure*}
\includegraphics[width=16cm]{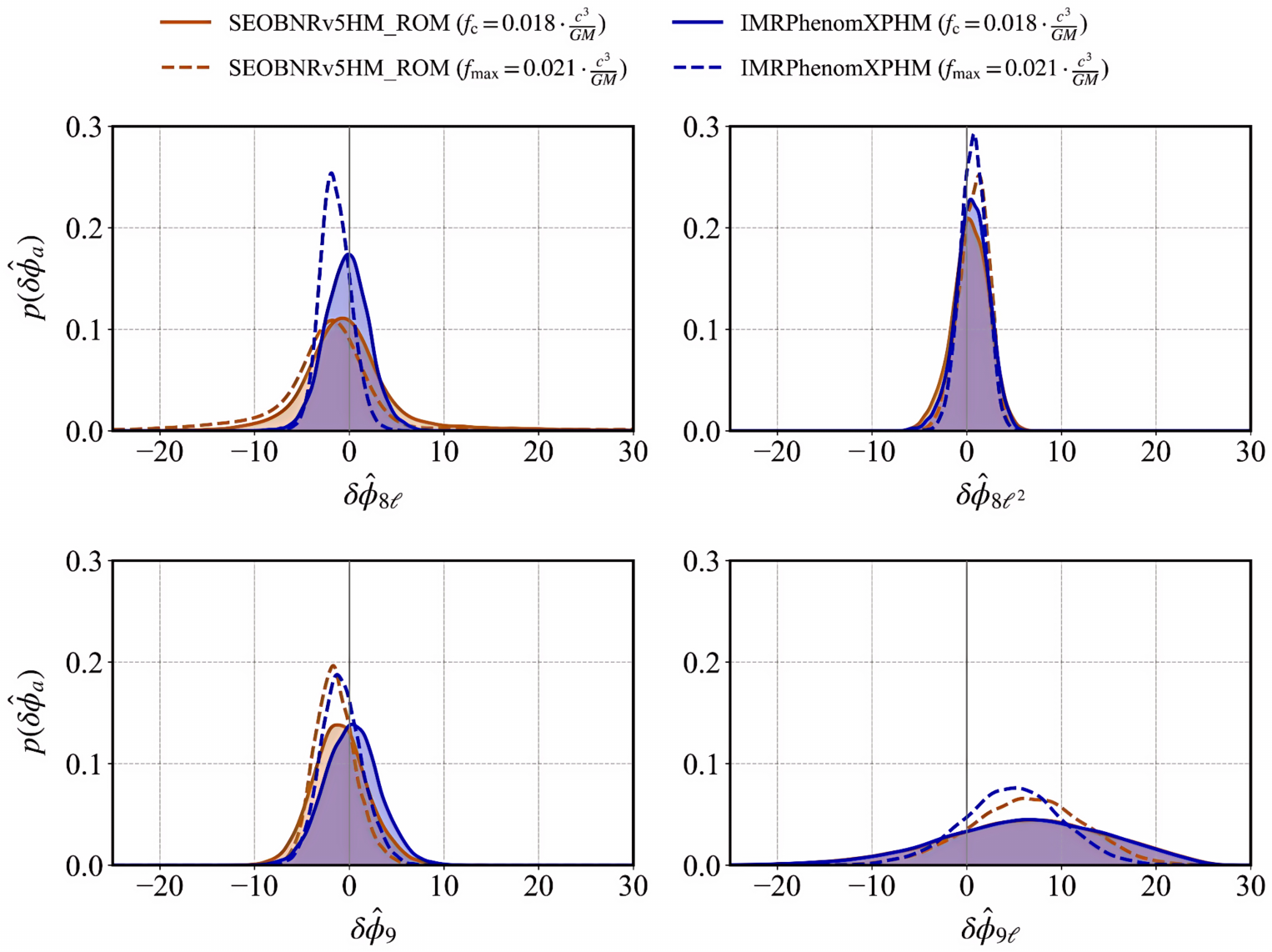}
\caption{Posterior distributions of the deviation parameters $\delta\hat{\phi}_a$ for GW230627 obtained by using two differenet cutoff frequencies $f_c = 0.018\,c^3/(GM)$  and $f_{\rm max} = 0.021\,c^3/(GM)$.}
\label{fig:phi_bounds4}
\end{figure*}

Figures~\ref{fig:phi_bounds3} and~\ref{fig:phi_bounds4} compare the posterior distributions of the four deviation parameters obtained with $f_c=0.018\,c^3/(GM)$ and $f_{\max} = 0.021\,c^3/(GM)$ for two GW events, GW250114 and GW230627, respectively. Although the posteriors obtained with the two cutoff frequencies are not identical, they are broadly consistent with each other. In particular, the posterior shapes and the overall agreement with GR remain unchanged. The results obtained with $f_{\max} = 0.021\,c^3/(GM)$ are slightly tighter than those obtained with $f_c = 0.018\,c^3/(GM)$, as expected, since the higher cutoff frequency includes a longer portion of the inspiral signal and therefore provides more information for constraining deviations from GR.

Overall, this comparison indicates that our main conclusions are robust against reasonable changes in the inspiral cutoff frequency.

\end{document}